\documentclass[11pt,a4paper]{article}
\usepackage[utf8]{inputenc}
\usepackage[margin=25mm]{geometry}
\usepackage{amsmath,amssymb,graphicx,booktabs}
\usepackage[super,sort&compress]{natbib}
\setcitestyle{comma}

\usepackage[font=small,labelfont=bf,labelsep=period]{caption}
\usepackage[hidelinks]{hyperref}
\graphicspath{{Figure/}}
\hypersetup{pdftitle={Light-induced interlayer spacing dynamics via orbital phonon coupling},pdfauthor={Wenwen Mao, Hanbyul Kim, Angel Rubio, Dongbin Shin}}
\begin{document}
\begin{center}
{\LARGE\bfseries Light-induced interlayer spacing dynamics via orbital phonon coupling\par}
\vspace{1em}
{\large Wenwen Mao$^{1,\dagger}$, Hanbyul Kim$^{1,\dagger}$,
Angel Rubio$^{2,3,4}$ and Dongbin Shin$^{1,2,*}$\par}
\vspace{0.7em}
{\small
$^1$Department of Physics and Photon Science, Gwangju Institute of Science and Technology (GIST), Gwangju 61005, Republic of Korea\par
$^2$Max Planck Institute for the Structure and Dynamics of Matter, Luruper Chaussee 149, 22761 Hamburg, Germany\par
$^3$Nano-Bio Spectroscopy Group, Departamento de Fisica de Materiales, Universidad del Pa\'is Vasco UPV/EHU, 20018 San Sebasti\'an, Spain\par
$^4$Initiative for Computational Catalysis (ICC), The Flatiron Institute, 162 Fifth Avenue, New York, NY 10010, USA\par
\vspace{0.5em}
$^\dagger$Wenwen Mao and Hanbyul Kim contributed equally to this work.\par
$^*$Correspondence: \href{mailto:dshin@gist.ac.kr}{dshin@gist.ac.kr}\par}
\end{center}

\section*{Abstract}
Interlayer coupling controls the electronic properties of layered van der Waals transition metal dichalcogenides. 
We investigate light-induced control of the interlayer spacing through orbital-selective excitation in trilayer 1T$^{\prime}$-WSe$_2$ and 1T$^{\prime}$-WS$_2$. 
Real-time time-dependent density functional theory simulations show that the interlayer spacing contracts or expands depending on whether chalcogen p-orbital density is depleted from or accumulated in the interlayer region. 
Effective Lindblad models coupled to the lattice dynamics reproduce these contrasting responses with simplified dynamics. 
A single effective excited state captures the cosine-like displacive motion in trilayer 1T$^{\prime}$-WSe$_2$, whereas the shift of the equilibrium spacing in trilayer 1T$^{\prime}$-WS$_2$ requires two excited states with different electron–phonon couplings and relaxation channels. 
Static calculations at varied interlayer spacings indicate that these spacing changes modify the electronic gaps and could access different electronic phases. 
These results connect orbital redistribution, carrier relaxation, and interlayer breathing motion, and establish orbital-selective optical excitation as a route to tuning the electronic properties of layered materials.

\section*{Introduction}
Two-dimensional (2D) materials exhibit intriguing physical phenomena and a wide range of applications distinct from three-dimensional systems, primarily due to electron confinement~\cite{britnell_strong_2013,rivera_observation_2015,jindal_coupled_2023,qian_quantum_2014,xu_large-gap_2013,wu_prediction_2014, tang2017quantum}. 
For example, monolayer transition metal dichalcogenides (TMDs) exhibit strong exciton binding owing to reduced dielectric screening~\cite{rivera_observation_2015}.
In addition, monolayer 1T$^{\prime}$-TMDs provide a platform for the quantum spin Hall insulator (QSHI) phase~\cite {qian_quantum_2014,xu_large-gap_2013,wu_prediction_2014, tang2017quantum}.
Moreover, van der Waals (vdW) interactions between stacking layers allow vertical stacking of different 2D materials, forming vdW heterostructures that offer unprecedented opportunities to explore quantum electronics at the nanoscale~\cite{geim2013van}. 
For instance, bilayer T$_d$-MoTe$_2$ shows an in-plane metallic band structure and out-of-plane ferroelectricity with superconductivity~\cite{jindal_coupled_2023}.
In these vdW-layered systems, modifying the interlayer alignment angle can create moir\'e patterns and support emergent phenomena, such as superconductivity in twisted bilayer graphene and light-induced twisting in hetero-bilayer TMDs~\cite{cao_unconventional_2018,duncan_photoinduced_2025,jiang_twist_2026,Kapfer_twist_2023}. 
These properties suggest vdW-layered materials as platforms for quantum and topological devices, including electrical control of charge, Wigner crystals, the quantum Hall effect, and spin transport through helical edge states~\cite{wunderlich2010spin, pesin2012spintronics, mahfouzi2012spin, brune2012spin, qian_quantum_2014, mellnik2014spin, lian_topological_2018,Li_Wigner_2024,Serlin_QHE_2020}.

The displacive excitation of coherent phonons (DECP) mechanism~\cite{emeis2025coherent}, in which photoexcitation alters the electron density and initiates coherent phonon motions, plays a central role in ultrafast light-induced structural dynamics~\cite{zeiger1992theory, kuznetsov1994theory, dhar1994time, giret2011entropy, juraschek2018sum, lakehal2019microscopic, caruso2023quantum, vaswani_light-driven_2020,shin_light-induced_2024,Seo2026,heo_frustrated_2025,Sergej2026,forst_nonlinear_2011}.
Such structural distortions can drive light-induced changes in material properties in 2D layered systems~\cite{sie_ultrafast_2019,vaswani_light-driven_2020,Seo2026,Sergej2026}.
For example, applying a terahertz pulse to the Weyl semimetal $T_d$-WTe$_2$ induces an interlayer shear displacement that restores inversion symmetry and collapses Weyl points, resulting in a transition to a trivial phase~\cite{sie_ultrafast_2019}.
A recent study further demonstrated that a linearly polarized pump can induce transient magnetism in bilayer MoSe$_2$ via a beating of two light-driven phonons~\cite{Seo2026}.
These findings highlight that light-driven coherent phonons can be employed as a crucial link between lattice distortions and modification of material properties~\cite{de2021colloquium, juraschek2017ultrafast, mankowsky2017ultrafast, papalazarou2012coherent, sobota2014distinguishing,shin_light-induced_2024}.

Here, we compare two examples of light-induced interlayer dynamics.
Using real-time time-dependent density functional theory (rt-TDDFT) simulations~\cite{shin2016real}, we investigate how the ultrafast femtosecond laser pulse frequency affects interlayer spacing in Trilayer 1T$^\prime$-TMDs.
Our first-principles calculations reveal interlayer contraction in trilayer WSe$_2$ under $0.6$~eV excitation and expansion in trilayer WS$_2$ under $3.0$~eV excitation.
We show that each structural response arises from changes in chalcogen $ p$-orbital occupation.
We employ effective Lindblad models to examine how excited-state populations and state-dependent electron--phonon couplings shape the interlayer breathing motion.
Finally, we compare the electronic consequences of contraction and expansion using static calculations at varied interlayer spacings.
This common sequence connects orbital redistribution, lattice dynamics, and electronic structure in the two materials.

\begin{figure}[!t]
    \centering
\includegraphics[width=\linewidth,height=0.60\textwidth,keepaspectratio]{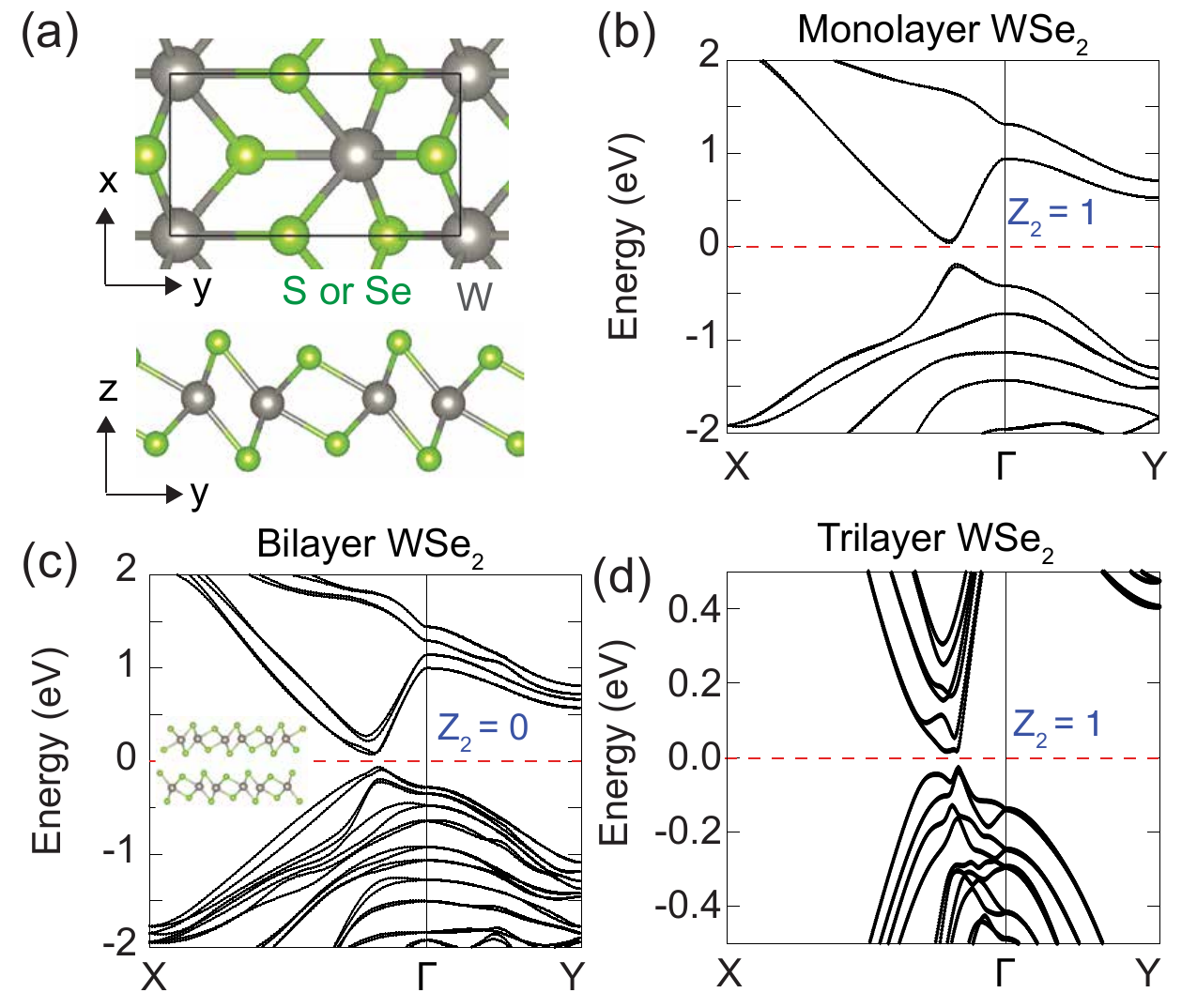}
\caption{\label{fig1}
\textbf{Atomic structure and layer-dependent electronic band structures of 1T$^\prime$-WSe$_2$.}
\textbf{a}, Atomic geometry of monolayer 1T$^{\prime}$-WSe$_2$.
\textbf{b--d}, Electronic band structures of (b) monolayer, (c) bilayer with AB stacking, and (d) trilayer 1T$^\prime$-WSe$_2$ with ABA stacking evaluated by the HSE06 hybrid functional. The inset of (c) is the atomic geometry of bilayer 1T$^{\prime}$-WSe$_2$ with AB stacking.}
\end{figure}

\section*{Results}

\subsection*{Layer-dependent electronic structure of trilayer systems}

Layered TMDs provide a platform to examine how interlayer coupling modifies the electronic structure through stacking~\cite{qian_quantum_2014,tang2017quantum}.
Layered TMDs, including MoX$_2$ and WX$_2$ ($X=\mathrm{S}$, Se, or Te), exhibit various electronic properties governed by intralayer coordination and interlayer stacking.
For example, monolayer MoS$_2$ in the trigonal-prismatic 1H structure, commonly referred to as monolayer 1H-MoS$_2$, has each Mo atom coordinated by six S atoms, with the upper and lower S planes aligned in-plane.
Its direct band gap, broken inversion symmetry, and spin--orbit coupling make it a platform for valleytronics~\cite{mak_valley_2014}.
In contrast, the 1T structure has octahedral metal coordination, whereas the 1T$^\prime$ structure is a distorted derivative characterized by zigzag metal chains (see Fig.~\ref{fig1}(a))~\cite{qian_quantum_2014}.
In layered tellurides such as MoTe$_2$, the related T$_d$ structure differs from bulk 1T$^\prime$ primarily in its interlayer stacking.
These structural variations give rise to metallic, semimetallic, and narrow-gap electronic states~\cite{sie_ultrafast_2019}.
In particular, our calculations identify monolayer 1T$^\prime$-WSe$_2$ as a quantum spin Hall insulator (QSHI) with $\mathbb{Z}_2=1$ (see Fig.~\ref{fig1}(b))~\cite{qian_quantum_2014}.
When 1T$^\prime$-WSe$_2$ monolayers are stacked along the out-of-plane direction at the van der Waals separation, interlayer hybridization modifies the electronic structure and can alter the resulting topological phase. 
The bilayer has a finite band gap of approximately $88$~meV but becomes topologically trivial (see Fig.~\ref{fig1}(c)). 
In contrast, trilayer 1T$^\prime$-WSe$_2$ recovers non-trivial topology with a band gap of about $22$~meV. 
This alternating behavior is consistent with an odd--even layer-number dependence of the $\mathbb{Z}_2$ topology.
An even number of quantum spin Hall layers can pairwise hybridize their helical edge-state channels and become topologically trivial, whereas an odd number of layers leaves an unpaired state, preserving the nontrivial $\mathbb{Z}_2$ phase~\cite{li_stacking-dependent_2025,Martinez2015}.
In addition, monolayer 1T$^\prime$-WS$_2$ has a gap of $151$~meV and is classified as a QSHI, whereas trilayer 1T$^\prime$-WS$_2$ is a trivial semimetal.
These distinct ground states indicate that trilayer 1T$^\prime$-TMDs can be QSHI or trivial semimetals, depending on the stacking configuration and chalcogen species.
These properties make trilayer 1T$^\prime$-WX$_2$ a suitable platform for investigating band-gap modulation and topological phase transitions induced by external perturbations.
In contrast, bilayer 1T$^\prime$-WX$_2$ can show a band-gap modulation while retaining its topologically trivial character over the range of perturbations considered here.
In Supplementary Sections 1 and 2, we show the various stacking configurations and their electronic structures.

\begin{figure}[!t]
    \centering
\includegraphics[width=\linewidth,height=0.80\textwidth,keepaspectratio]{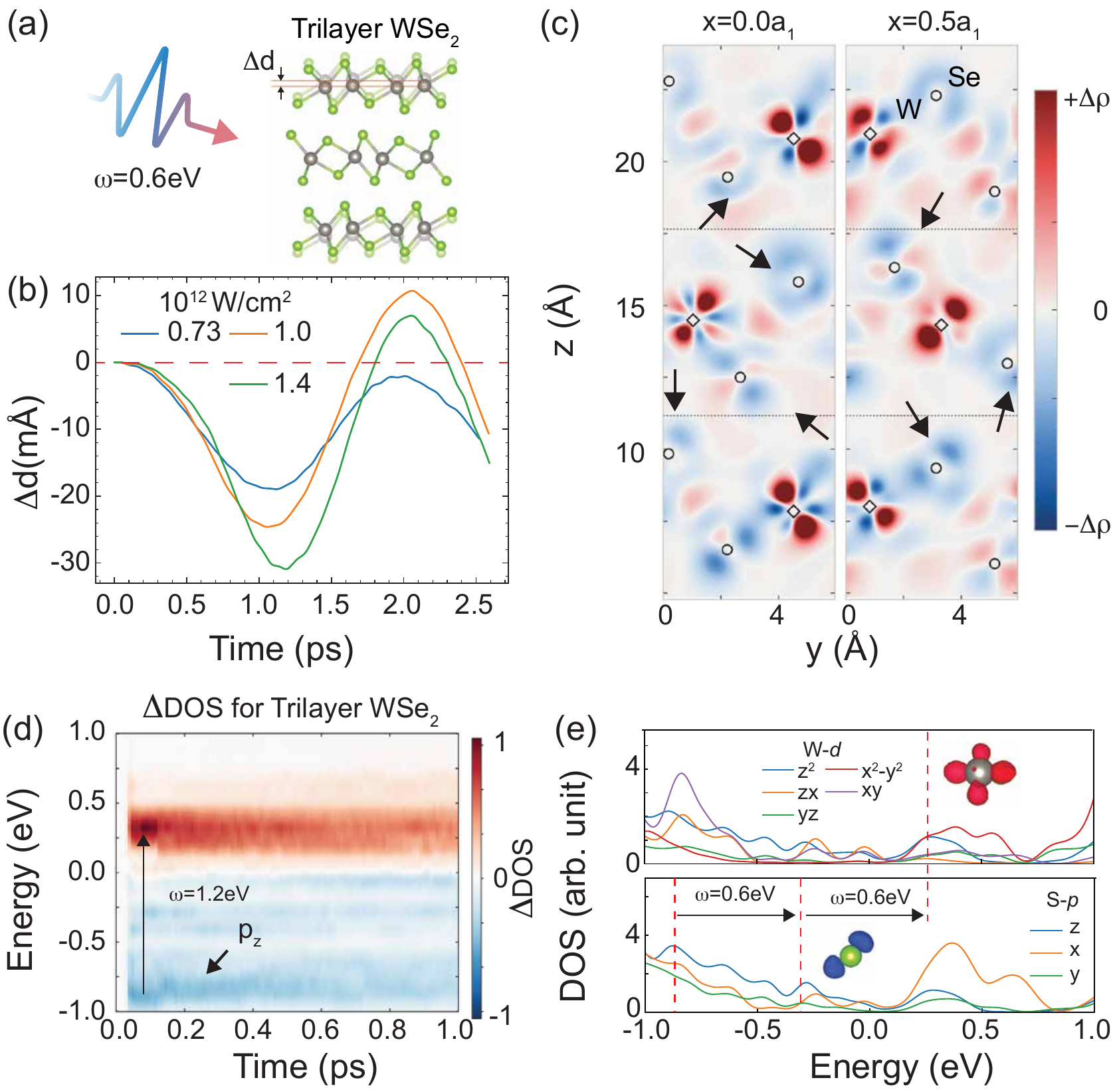}
\caption{\label{fig2}
\textbf{Light-induced interlayer contraction and electronic redistribution in trilayer 1T$^\prime$-WSe$_2$.}
\textbf{a}, Schematic of the interlayer contraction by light-induced breathing motion.
\textbf{b}, Light-induced interlayer-spacing change $\Delta d(t)=(d(t)-d(0))/2$ under $0.6$~eV excitation at different pump intensities, where $d$ is the outer-layer separation.
\textbf{c}, Time-averaged photo-induced electron-density difference $\Delta\rho$ in the $x=0.0a$ (left) and $x=0.5a$ (right) planes.
\textbf{d}, Energy- and time-resolved carrier redistribution from fixed-ion rt-TDDFT under $0.6$~eV excitation.
\textbf{e}, Orbital-projected density of states for W-$d$ and Se-$p$ orbitals.
In (c), open circle and open diamond indicate the location of W and Se atoms.
The inset of (e) indicates W-$d$ (upper) and Se-$p_z$ (lower) orbitals for electron and hole carriers, respectively.
}
\end{figure}

\subsection*{Interlayer contraction and a single-state response in trilayer 1T$^\prime$-WSe$_2$}

We first consider trilayer 1T$^\prime$-WSe$_2$, whose electronic structure is a QSHI with a $22$~meV indirect gap (Fig.~\ref{fig1}(d)).
To investigate the light-induced dynamics, we apply a y-polarized pulse with $\hbar\omega=0.6$~eV, $t_0=40$~fs, and $\sigma=10$~fs at peak intensities of $0.73, 1.0,$ and $1.4\times10^{12}$~W/cm$^2$ (Fig.~\ref{fig2}(a)).
By this ultrafast laser pump, the interlayer breathing motion is generated toward shorter separations, with a cosine-like displacement profile and a larger oscillation amplitude at higher intensity (Fig.~\ref{fig2}(b)).
Notably, the oscillation is asymmetric, with displacement away from the equilibrium position ($\Delta d=0$).
To understand the microscopic origin of this dynamics, we evaluate the charge-density difference between the ground state and the time-averaged state after laser pumping.
The density difference in the $x=0.0a$ and $x=0.5a$ planes shows electron depletion from Se-$p_z$ orbitals in the interlayer region and accumulation on W-$d$ orbitals (Fig.~\ref{fig2}(c)).
This charge-density change indicates that the modified charge density induced a modified potential on the layer breathing mode and launched motion via the DECP mechanism.
For trilayer 1T$^\prime$-WSe$_2$ under $0.6$~eV excitation, the energy- and time-resolved carrier distribution shows holes near $-0.9$~eV and electrons near $+0.3$~eV under two-photon excitation (Fig.~\ref{fig2}(d)).
Notably, the electron carrier near $+0.3$~eV remains within $1$~ps time propagation, while hole carriers are distributed over a wide energy range.
The projected density of states in Fig.~\ref{fig2}(e) estimates a possible photo-induced charge transfer from Se-$p_z$ to W-$d$ states between $-0.9$~eV and $+0.3$~eV.
The depletion of interlayer electron density provides a microscopic interpretation of the contraction as reduced interlayer electronic repulsion.

\begin{figure}[!t]
    \centering
\includegraphics[width=\linewidth,height=0.7\textwidth,keepaspectratio]{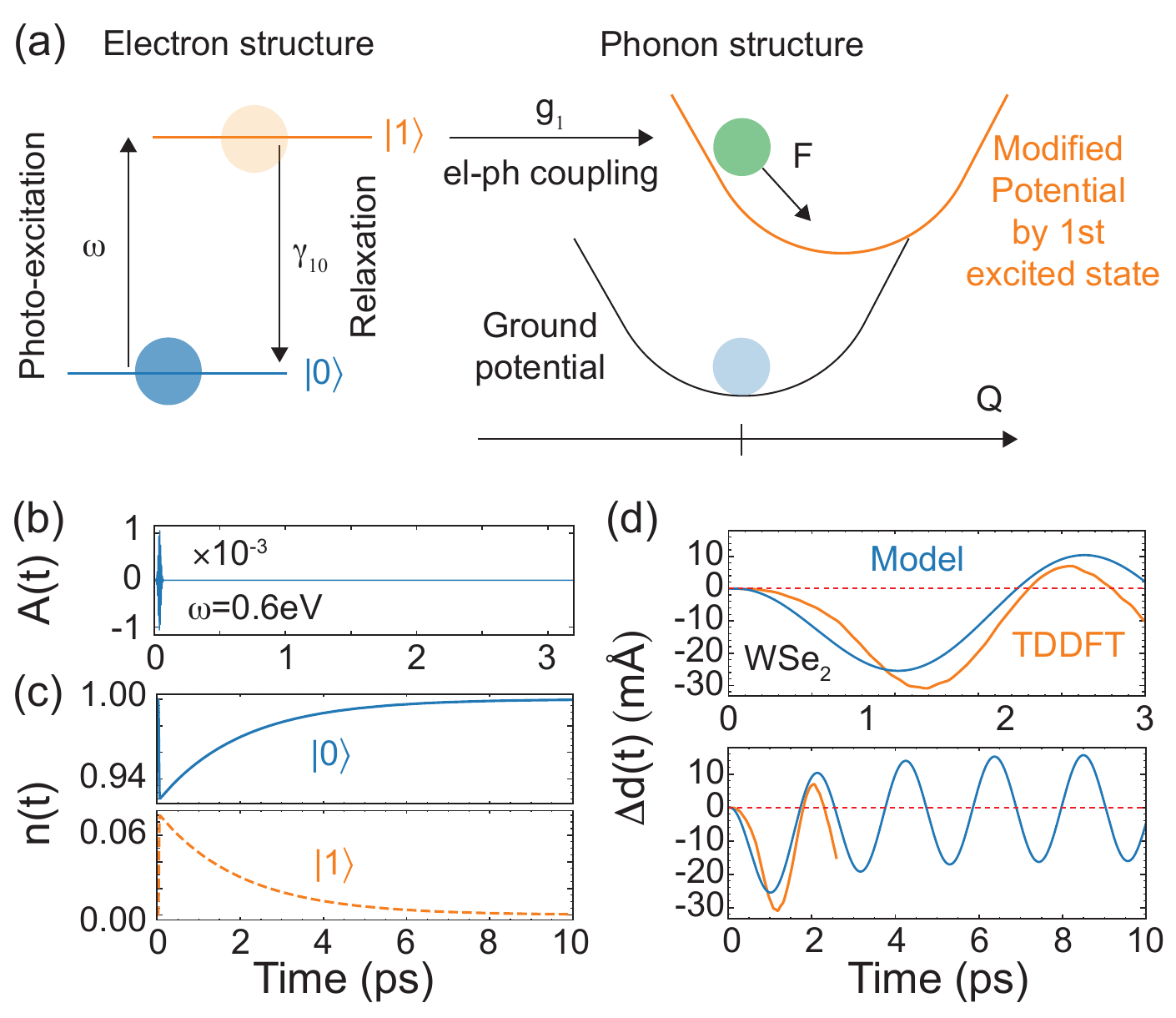}
\caption{\label{fig3}
\textbf{Two-level model of carrier populations and interlayer motion in trilayer 1T$^\prime$-WSe$_2$.}
\textbf{a}, Schematic of the two-level electronic model coupled to the interlayer breathing mode, with relaxation from $|1\rangle$ to $|0\rangle$.
\textbf{b}, Time profile of the optical coupling $A(t)$ with $\hbar\omega=0.6$~eV.
\textbf{c}, State populations $n_i(t)$ in the two-level model.
\textbf{d}, Interlayer-spacing change $\Delta d(t)$ from the Lindblad--lattice model and rt-TDDFT coupled to Ehrenfest dynamics, using $\Delta d(t)=(d(t)-d(0))/2$.}
\end{figure}

To understand how photoexcited carriers drive the interlayer-spacing changes observed in the rt-TDDFT simulations, we employ a minimal model that couples electronic dynamics to interlayer lattice motion.
For trilayer 1T$^\prime$-WSe$_2$ with a $0.6$~eV laser, the minimal model with a ground state $|0\rangle$ and one effective excited state $|1\rangle$, which is coupled to a phonon mode, is used (Fig.~\ref{fig3}(a)).
We solve the Lindblad master equation to simulate excited-state relaxation and evaluate the motion of the phonon with the achieved carrier occupation.
The effective excitation energy is set to $0.60$~eV, and the relaxation rate ($\gamma_{10}=0.5$~THz) is taken from the experimental report for the 1T$^\prime$-WTe$_2$ system ~\cite{zheng_quantum_2016}.
We extract the effective electron--phonon coupling parameter between the excited state and breathing mode ($g_1=5.2\times10^{-3}$~Ry/bohr) and the breathing-mode frequency ($\omega_{\mathrm{ph}}=0.47$~THz) by fitting the rt-TDDFT results.
When laser pulse with $A_0=1.08\times10^{-3}$~Ry amplitude is applied, a peak excited-state population of $n_1=0.075$ is found (Figs.~\ref{fig3}(b) and~\ref{fig3}(c)).
With a positive electron--phonon coupling parameter ($g_1>0$), the excited population induces a contraction-directed force on the breathing phonon mode.
As a result, this two-level model reproduces the cosine-like interlayer response of the rt-TDDFT trajectory and provides a single-state, DECP-type description (Fig.~\ref{fig3}(d))~\cite{emeis2025coherent}.
This indicates that a simple two-level model with strong electron-phonon coupling can reproduce the light-induced dynamics for trilayer 1T$^\prime$-WSe$_2$ with a $0.6$~eV laser.

\begin{figure}[!t]
    \centering
\includegraphics[width=\linewidth,height=0.80\textwidth,keepaspectratio]{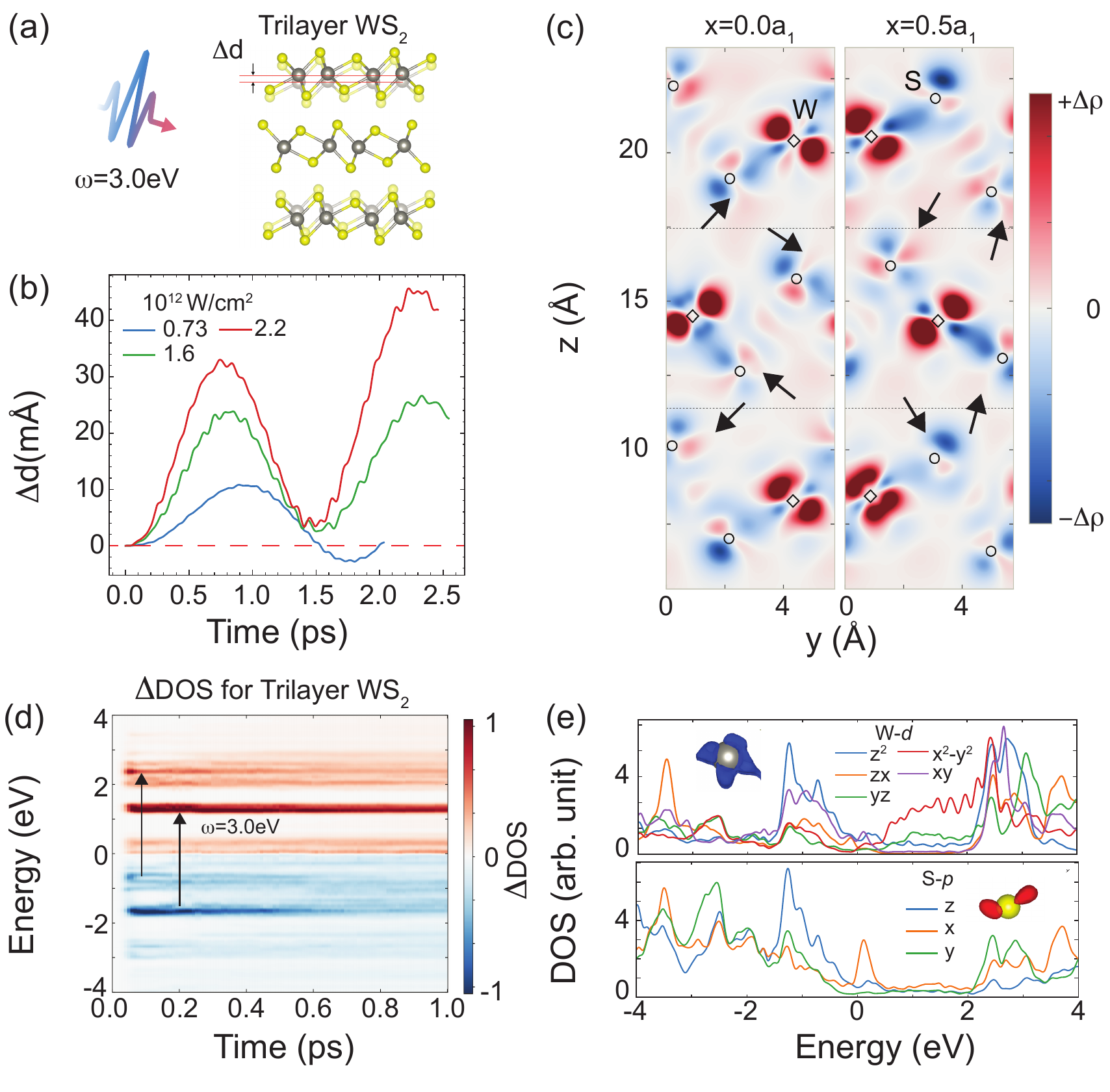}
\caption{\label{fig4}
\textbf{Light-induced interlayer expansion and electronic redistribution in trilayer 1T$^\prime$-WS$_2$.}
\textbf{a}, Schematic of the interlayer expansion by light-induced breathing motion.
\textbf{b}, Interlayer-spacing change $\Delta d(t)=(d(t)-d(0))/2$ under $3.0$~eV excitation at different pump intensities, where $d$ is the outer-layer separation.
\textbf{c}, Time-averaged photo-induced electron-density difference $\Delta\rho$ in the $x=0.0a$ (left) and $x=0.5a$ (right) planes.
\textbf{d}, Energy- and time-resolved carrier redistribution from fixed-ion rt-TDDFT under $3.0$~eV excitation.
\textbf{e}, Orbital-projected density of states for W-$d$ and S-$p$ orbitals.
In (c), open circle and open diamond indicate the location of W and S atoms.
The inset of (e) indicates W-$d$ (upper) and S-$p_y$ (lower) orbitals for hole and electron carriers, respectively.
}
\end{figure}

\subsection*{Interlayer expansion and sequential relaxation in 1T$^\prime$-\texorpdfstring{WS$_2$}{WS2}}

Trilayer 1T$^\prime$-WS$_2$ is a trivial semimetal with no insulating band gap in its relaxed geometry.
In this case, we apply y-polarized pulses with $\hbar\omega=3.0$~eV, $t_0=50$~fs, and $\sigma=20$~fs at different peak intensities of $0.73, 1.6,$ and $2.2\times10^{12}$~W/cm$^2$ (Fig.~\ref{fig4}(a)).
The rt-TDDFT Ehrenfest dynamics simulation shows that the resulting interlayer breathing motion is directed toward larger separations, corresponding to positive $\Delta d$ (Fig.~\ref{fig4}(b)).
Notably, the equilibrium shifts significantly away from the original point, especially at $2.2\times10^{12}$~W/cm$^2$ intensity, unlike a simple cosine oscillation following the DECP mechanism.
Using the same density-difference analysis, we find electron accumulation on S-$p_y$ orbitals in the interlayer region and depletion from W-$d$ orbitals in the $x=0.0a$ and $x=0.5a$ planes (Fig.~\ref{fig4}(c)).
We also evaluate the energy- and time-resolved carrier distribution to understand the microscopic origin of this behavior, as in the previous case (Fig.~\ref{fig4}(d)).
It shows that hole carriers emerge at $-1.68$~eV and $-0.61$~eV, and electron carriers appear at $1.26$~eV and $2.24$~eV under a $3.0$~eV laser.
Notably, long-lived carriers at $1.26$ and $-1.68$~eV are found within $1$ ps, while carriers at other energy levels spread quickly.
These two excitation channels are distinguishable from the previous case.
The orbital-resolved density of states associates this possible photo-excitation between W-$d_{zx}$/$d_{yz}$ and S-$p_y$ states (Inset of Fig.~\ref{fig4}(e)).
The accumulation of electron density at the interlayer region expands the interlayer distance due to increased interlayer electronic repulsion.
Thus, a $3.0$~eV laser on trilayer 1T$^\prime$-WS$_2$ induces different orbital occupations, relaxation, and structural displacements than a $0.6$~eV laser on Trilayer 1T$^\prime$-WSe$_2$.
From Bader charge analysis, we found that the central layer loses about $0.005$~$e$/cell to the outer layers in both cases (WSe2: $0.6$~eV, $1.4 \times 10^{12}W/cm^{2}$; WS2: $3.0$~eV, $2.2 \times 10^{12}W/cm^{2}$).
This shows that light-modified orbital occupation at the chalcogen site can control interlayer spacing rather than charge redistribution between layers.
In Supplementary Section 3, we provide details of the laser-pulse frequency dependence of the interlayer spacings.

\begin{figure}[!t]
    \centering
\includegraphics[width=\linewidth,height=0.70\textwidth,keepaspectratio]{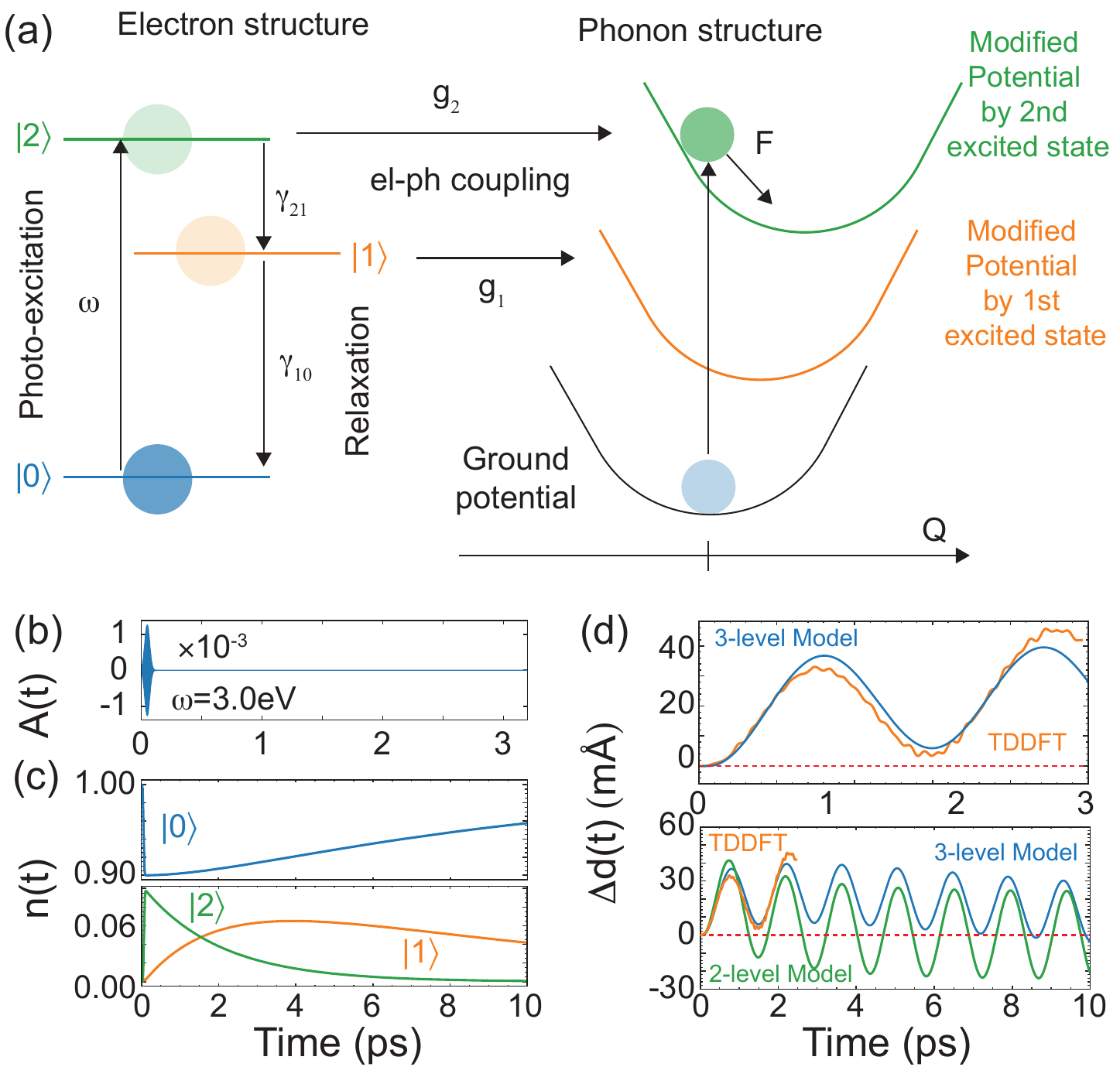}
\caption{\label{fig5}
\textbf{Carrier redistribution and a three-level model of interlayer motion in trilayer 1T$^\prime$-WS$_2$.}
\textbf{a}, Three-level electronic model coupled to the interlayer breathing mode, with relaxation from $|2\rangle$ to $|1\rangle$ and from $|1\rangle$ to $|0\rangle$.
\textbf{b}, Time profile of the model optical coupling $A(t)$ for $\hbar\omega=3.0$~eV.
\textbf{c}, State populations $n_i(t)$ in the three-level model.
\textbf{d}, Interlayer-spacing change $\Delta d(t)$ from the Lindblad--lattice model and rt-TDDFT coupled to Ehrenfest dynamics.}
\end{figure}

To understand this difference from the former case, we construct a simple model for trilayer 1T$^\prime$-WS$_2$ with a $3.0$~eV laser.
The two excitation channels are described by the three-level carrier dynamics with a ground state $|0\rangle$ and two effective excited states, $|1\rangle$ with slow decay and $|2\rangle$ with fast decay (Fig.~\ref{fig5}(a)).
The effective excitation energies are set as $\epsilon_1=2.9$~eV and $\epsilon_2=3.0$~eV, with relaxation rates $\gamma_{21}=0.5$~THz and $\gamma_{10}=0.111$~THz achieved from experimental report~\cite{zheng_quantum_2016}.
We extracted the coupling strengths $g_1=-1.2\times10^{-2}$~Ry/bohr and $g_2=-6.3\times10^{-3}$~Ry/bohr, breathing-mode frequency $\omega_{\mathrm{ph}}=0.70$~THz, and pulse amplitude $6.3\times10^{-4}$~Ry by fitting to rt-TDDFT simulation results with a $2.2\times10^{12}$~W/cm$^2$ laser pump.
This optical pulse generates excited-state populations, which relax through the $|2\rangle\rightarrow|1\rangle$ and $|1\rangle\rightarrow|0\rangle$ channels (Figs.~\ref{fig5}(b) and~\ref{fig5}(c)).
The two negative coupling strengths provide an expansion-directed force.
In contrast, a two-level model with $g=-9.4\times 10^{-3}$~Ry/bohr fails to describe this displaced equilibrium behavior (Fig.~\ref{fig5}(d)).
This result indicates that this different behavior originates from two electron-phonon coupling channels with different decay rates.

Taken together, the two cases support a common relation between the X-$p$ orbital occupations and the direction of the interlayer response.
The comparison identifies two distinct effective forces on the interlayer spacing within a common population-driven framework.
A single positive coupling produces the WSe$_2$ contractive oscillation via the unoccupied Se-$p_z$ orbital.
In contrast, two negative couplings with different carrier decay rates introduce the WS$_2$ interlayer expansion with displaced equilibrium via the occupied S-$p_y$ orbital.
These results indicate that light-driven interlayer spacing arises from complex carrier dynamics that depend on the occupation configuration and decay channel of excited carriers.

\begin{figure}[!t]
    \centering
\includegraphics[width=\linewidth,height=0.60\textwidth,keepaspectratio]{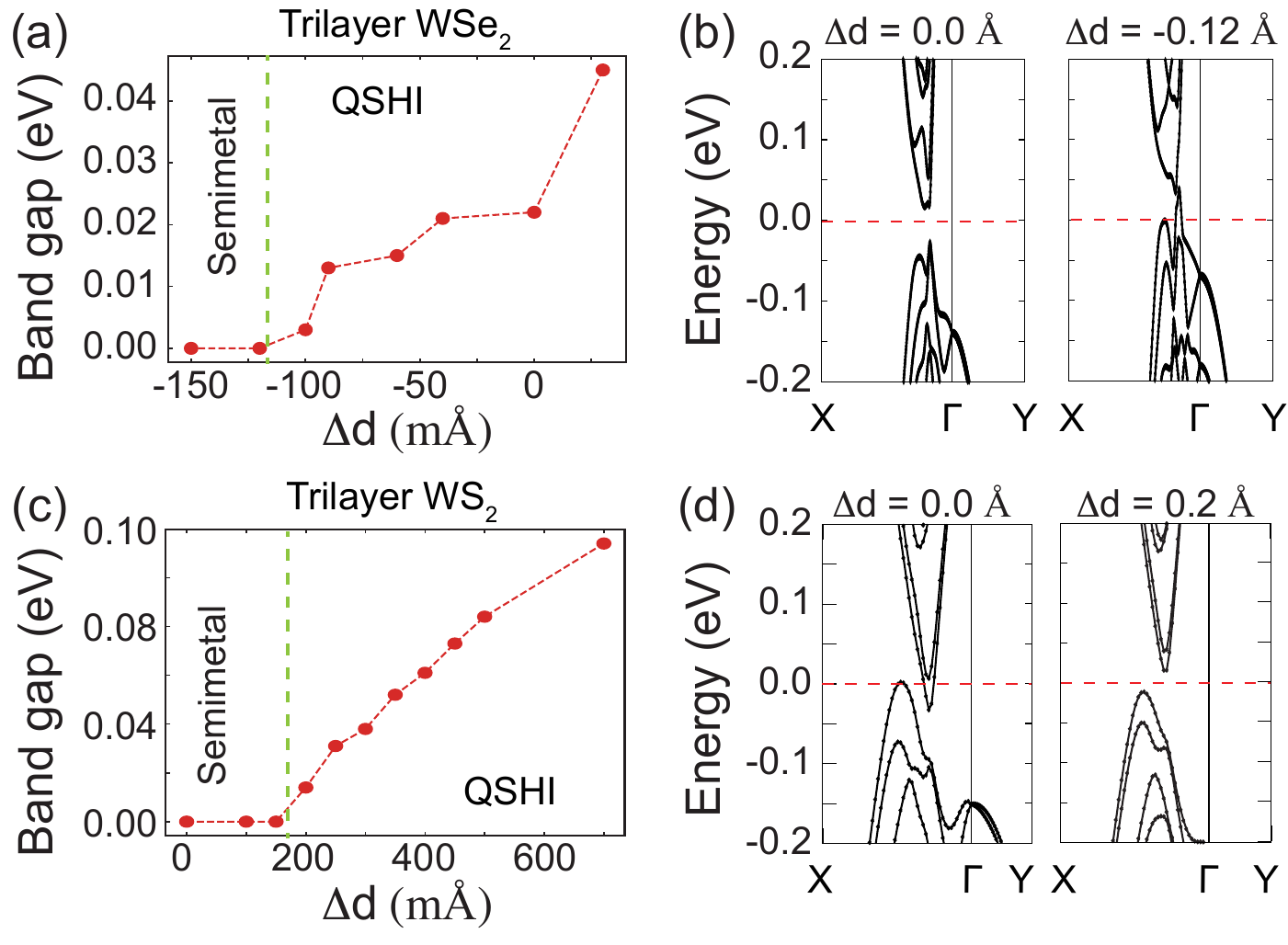}
\caption{\label{fig6}
\textbf{Spacing-dependent electronic structure of trilayer 1T$^\prime$-WSe$_2$ and 1T$^\prime$-WS$_2$.}
\textbf{a}, Band gap and reported $Z_2$ characterization of trilayer 1T$^\prime$-WSe$_2$ as a function of $\Delta d$ from static calculations.
\textbf{b}, Trilayer 1T$^\prime$-WSe$_2$ band structures at the relaxed geometry (left) and at $\Delta d=-0.12$~\AA\ (right).
\textbf{c}, Band gap and reported $Z_2$ characterization of WS$_2$ as a function of $\Delta d$ from static calculations.
\textbf{d}, Trilayer 1T$^\prime$-WS$_2$ band structures at the relaxed geometry (left) and at $\Delta d=0.20$~\AA\ (right).
Here $\Delta d=(d-d_0)/2$ uses the same spacing convention as the dynamical comparisons, with $d_0$ the relaxed outer-layer separation.}
\end{figure}

\subsection*{Modulation of electronic structure via interlayer contraction and expansion}

This light-modified interlayer spacing in trilayer 1T$^\prime$-WX$_2$ modifies the electronic structure.
We evaluate the indirect band gap and topological characterization at statically varied interlayer spacings using HSE hybrid-functional calculations for an accurate description.
For the trilayer 1T$^\prime$-WSe$_2$ case, reducing the interlayer spacing narrows the band gap and introduces a topological phase transition to a semimetallic phase from a QSHI for $\Delta d<-0.12$~\AA\ (Fig.~\ref{fig6}(a)).
The band structures show gap closure near the X--$\Gamma$ line, consistent with increased band dispersion from enhanced interlayer coupling (Fig.~\ref{fig6}(b)).
Therefore, interlayer distance contraction in trilayer 1T$^\prime$-WSe$_2$ reduces the band gap and can drive a topological phase transition toward the trivial semimetal.
For the trilayer 1T$^\prime$-WS$_2$ case, increasing the interlayer spacing opens a band gap and induces a QSHI phase reported for $\Delta d>0.18$~\AA\ (Fig.~\ref{fig6}(c)).
The changes near the X--$\Gamma$ line are consistent with gap opening as interlayer coupling is reduced (Fig.~\ref{fig6}(d)).
This result indicates that interlayer distance expansion in trilayer 1T$^\prime$-WS$_2$ opens the band gap and can lead to a topological phase transition toward a QSHI.
Although our rt-TDDFT simulation does not provide enough interlayer spacing to compare with our static analysis of the topological phase transition, our first-principles calculation result may under- or overestimate the critical distance due to DFT inaccuracies in the band gap.
Experiments can observe our theoretical prediction using time-resolved X-ray diffraction for the light-induced breathing mode and time- and angle-resolved photoemission spectroscopy for ultrafast carrier dynamics~\cite{heo_frustrated_2025}.
In Supplementary Section 4, we provide details of the electronic-structure analysis at various interlayer spacings.

\section*{Discussion}

We studied light-induced interlayer spacing dynamics in trilayer 1T$^{\prime}$-TMDs using rt-TDDFT calculations coupled to Ehrenfest ionic dynamics~\cite{shin2016real}, complemented by effective Lindblad--lattice models and static hybrid-functional calculations.
We show that the orbital character of the photoexcited carriers can selectively control the direction of interlayer breathing in trilayer 1T$^\prime$-WSe$_2$ and 1T$^\prime$-WS$_2$.
The optical excitation redistributes orbitals between W-$d$ and X-$p$ states.
These excitations modify the interlayer spacing in opposite directions, depending on which orbitals the pump populates via the DECP mechanism~\cite{zeiger1992theory,emeis2025coherent}.
When $0.6$~eV light is applied to the trilayer 1T$^\prime$-WSe$_2$, electrons transition from Se-$p_z$ states to W-$d$ states, inducing layer contraction.
This light-induced dynamics shows a cosine-like DECP-type displacement whose amplitude grows with the pump intensity.
When $3.0$~eV light is applied to the trilayer 1T$^\prime$-WS$_2$, electrons transition from W-$d$ states to S-$p_y$ states, leading to layer expansion.

We also set up a simple model and investigate it using the Lindblad master equation to understand how carrier dynamics affect interlayer breathing motion.
For trilayer 1T$^\prime$-WSe$_2$ with $0.6$~eV light, a simple two-level model reproduced a contraction-directed oscillation of interlayer breathing motion with simple cosine-like DECP motion.
In contrast, a three-level model with different decay rates for two excited channels is required to describe the light-induced dynamics in trilayer 1T$^\prime$-WS$_2$ with $3.0$~eV light, which shows a shifted equilibrium oscillation position.
This result indicates that complex carrier dynamics can produce oscillations with lattice distortion beyond simple DECP behavior.

Our static electronic structure analysis suggests possible experimental observables.
Static DFT calculation results show that contraction closes the $22$~meV gap of trilayer 1T$^\prime$-WSe$_2$ and drives it into a semimetallic regime for $\Delta d<-0.12$~\AA, whereas expansion opens a gap in trilayer 1T$^\prime$-WS$_2$ and turns it into a QSHI for $\Delta d>0.18$~\AA.
This result indicates that orbital-selective excitation, carrier evolution, and interlayer motion together provide a versatile framework for manipulating the electronic properties of layered TMDs, with the photon energy and the pump intensity serving as effective tuning parameters.

\section*{Methods}
\subsection*{Density functional theory calculation}
We relaxed the ground-state structures of monolayer, bilayer, and trilayer 1T$^\prime$-WX$_2$ ($X=\mathrm{S}$ or Se) using density functional theory as implemented in Quantum ESPRESSO~\cite{giannozzi2009quantum}.
The structures were optimized until the atomic forces were below $10^{-5}$~eV/\AA\ and the total-energy change was below $10^{-11}$~eV.
We used norm-conserving pseudopotentials including spin-orbit coupling (SOC) and the Perdew--Burke--Ernzerhof (PBE) exchange-correlation functional~\cite{perdew1996generalized}.
Interlayer vdW interactions were treated using Grimme's DFT-D3 correction with Becke--Johnson damping (DFT-D3BJ)~\cite{grimme2010consistent,grimme2011effect}.
We used a $6\times6\times1$ Monkhorst--Pack grid for Brillouin-zone sampling and a $15$~\AA\ vacuum region to reduce interactions under periodic boundary conditions.

For the electronic-structure and topology analysis, we used the HSE06 hybrid functional~\cite{heyd2003hybrid} for monolayer, bilayer, and trilayer 1T$^\prime$-WX$_2$.
We used the atomic structures achieved from the PBE-relaxed geometry.
These calculations included SOC and the vdW correction described above and used a plane-wave energy cutoff of $80$~Ry.
We constructed maximally localized Wannier functions using Wannier90~\cite{mostofi2008wannier90} and evaluated the band structures and $Z_2$ topological characterization using WannierTools~\cite{wu2018wanniertools}.

\subsection*{TDDFT calculation} 
We performed Ehrenfest dynamics using a homemade rt-TDDFT code based on Quantum ESPRESSO and tested it with Octopus~\cite{shin2016real,tancogne-dejean_octopus_2020}.
We considered a $6\times6\times1$ Brillouin-zone sampling and a plane-wave energy cutoff of $50$~Ry.
We treated interlayer vdW interactions with the same DFT-D3BJ correction used in the ground-state calculations~\cite{grimme2010consistent,grimme2011effect}.
A Crank--Nicolson-type propagator was used with a time step of $2.4$~as. For both materials, we applied an electric-field pulse
\begin{equation}
E(t)=E_0\cos(\omega t)\exp\!\left[-\frac{(t-t_0)^2}{2\sigma^2}\right],
\end{equation}
where $\hbar\omega$ is the photon energy, $t_0$ is the envelope center, and $\sigma$ is the standard deviation of the field envelope.
We define the structural observable as $\Delta d(t)=(d(t)-d(0))/2$, where $d(t)$ is the distance between the outer layers: $\Delta d<0$ denotes contraction and $\Delta d>0$ denotes expansion.
To characterize the light-induced charge redistribution, we evaluated $\Delta\rho(\mathbf r)=\bar\rho^{\mathrm{exci}}(\mathbf r)-\rho^0(\mathbf r)$, the difference between the time-averaged excited-state and ground-state electron densities.
Positive and negative $\Delta\rho$ indicate electron accumulation and depletion, respectively.

We also evaluate energy- and time-resolved carrier distributions.
For fixed-ion TDDFT trajectories, a projected occupation of the ground-state Kohn--Sham ($\psi_{i\mathbf k}(0)$) is defined as follows:
\begin{equation}
N_{i\mathbf k}(t)=\sum_j f_{j\mathbf k}
\left|\left\langle\psi_{i\mathbf k}(0)\middle|\psi_{j\mathbf k}(t)\right\rangle\right|^2,
\label{eq:occupation}
\end{equation}
where $f_{j\mathbf k}$ is the occupation of initial propagating state ($\psi_{j\mathbf k}(t)$).
The corresponding energy-resolved occupation change is
\begin{equation}
\Delta D(E,t)=\sum_{i\mathbf k} w_{\mathbf k}
\left[N_{i\mathbf k}(t)-f_{i\mathbf k}\right]
\delta(E-\epsilon_{i\mathbf k}),
\label{eq:carrier}
\end{equation}
where $w_{\mathbf k}$ is the Brillouin-zone weight and $\epsilon_{i\mathbf k}$ is the ground-state energy.
Positive and negative values represent electron and hole carriers, respectively.

\subsection*{Lindblad master equation}
We describe effective carrier relaxation with a Lindblad master equation for the electronic density matrix $\rho$,
\begin{equation}
\dot\rho=-\frac{i}{\hbar}[H(t),\rho]
+\sum_{i\to j}\gamma_{ij}\left(
L_{ij}\rho L_{ij}^{\dagger}
-\frac{1}{2}\left\{L_{ij}^{\dagger}L_{ij},\rho\right\}
\right),
\label{eq:lindblad}
\end{equation}
where $L_{ij}=|j\rangle\langle i|$, $\{A,B\}=AB+BA$, and $\gamma_{ij}$ is the population-transfer rate from state $i$ to state $j$, respectively.
The model contains a ground state $|0\rangle$ and one excited state for WSe$_2$ or two excited states for WS$_2$.
The WSe$_2$ model contains the $1\to0$ relaxation channel; the WS$_2$ model contains the $2\to1$ and $1\to0$ channels.
Taking the ground-state energy as zero, the electronic Hamiltonian is
\begin{equation}
H(t)=\sum_{i=1}^{N_{\mathrm{ex}}}\epsilon_i|i\rangle\langle i|
+A(t)\sum_{j=1}^{N_{\mathrm{ex}}}
\left(|0\rangle\langle j|+|j\rangle\langle0|\right),
\end{equation}
with
\begin{equation}
A(t)=A_0\cos(\omega t)\exp\!\left[-\frac{(t-t_0)^2}{2\sigma^2}\right].
\end{equation}
Here $A(t)$ denotes an effective optical coupling energy.
The rates are $\gamma_{10}=0.5$~THz for WSe$_2$ and $\gamma_{21}=0.5$~THz and $\gamma_{10}=0.111$~THz for WS$_2$.
This carrier relaxation rate in the TMD system follows the previous experimental report for $1T^\prime$-WTe$_2$~\cite{zheng_quantum_2016}.
We obtain the breathing-mode frequency, damping, and coupling constants by least-squares fitting of the model $\Delta d(t)$ to the TDDFT trajectories.
The lattice equation of motion is given as follows: $M \ddot{Q} = -M \omega^2 Q - \sum_i g{_i} n{_i}(t)$, where $M$ and $Q$ are the effective phonon mass and normal coordinate, respectively.
We evaluate the ionic trajectory using excited-state occupation from the Lindblad master equation.

\section*{Data availability}
The data that support the findings of this article are openly available at figshare

\section*{Code availability}
The in-house rt-TDDFT code and Lindblad--lattice modeling scripts used in this study are available from the corresponding authors on reasonable request.

\section*{Acknowledgments}
\subsection*{Funding}
We acknowledge the European Research Council, ERC-2024-SyG-UnMySt-101167294).
We acknowledge the Deutsche Forschungsgemeinschaft, EXC 2056 - project ID 390715994, Grupos Consolidados (IT1249-19), and SFB925 "Light induced dynamics and control of correlated quantum systems". 
D.S. was supported by the National Research Foundation of Korea (NRF) grant funded by the Korea government (MSIT) (No. RS-2026-25508298).
This research was supported by the ANCHOR program through the Gwangju ANCHOR Center, funded by the Ministry of Education(MOE) and the Jeonnam-Gwangju Special Metropolitan City, Republic of Korea. (2026-ANCHOR-05-001).
We acknowledge the National Supercomputing Center with supercomputing resources, including technical support, KSC-2025-CRE-0001. 

\subsection*{Use of AI-assisted tools}
During the preparation of this work, the authors used Grammarly and Claude to improve readability and language and to format the text according to the journal template.

\section*{Competing interests}
The authors declare no conflicts of interest.

\section*{Author Information}
\subsection*{Contributions}
 All authors wrote the manuscript. 
W.M. and H.K. performed the ab initio calculations under the supervision of A.R. and D.S.. 
All authors discussed the results and contributed to the final paper.

\subsection*{Corresponding authors}
Correspondence to Dongbin Shin.

\end{document}